# On the density of primes of the form $\boldsymbol{X^2 + c}$

<u>4th November 2023</u>

**WOLF Marc,** https://orcid.org/0000-0002-6518-9882

Email: marc.wolf3@wanadoo.fr

**WOLF François,** https://orcid.org/0000-0002-3330-6087

Email: francois.wolf@dbmail.com

**Abstract**

We present a method for finding large fixed-size primes of the form $X^2 + c$. We study the density of primes on the sets $E_c = \{N(X, c) = X^2 + c, X \in (2\mathbb{Z} + (c-1))\}$, $c \in \mathbb{N}^*$. We describe an algorithm for generating values of $c$ such that a given prime $p$ is the minimum of the union of prime divisors of all elements in $E_c$. We also present quadratic forms generating divisors of $E_c$ and study the prime divisors of its terms. This paper uses the results of Dirichlet's arithmetic progression theorem [1] and the article [6] to rewrite a conjecture of Shanks [2] on the density of primes in $E_c$. Finally, based on these results, we discuss the heuristics of large primes occurrences in the research set of our algorithm.

**<u>Keywords:</u>** density of primes, large primes, quadratic form, arithmetic sequence, constant density.

# SOMMAIRE

# 1. Introduction

## 1.1. Density of primes

The prime number theorem states that prime numbers tend to get scarcer as they increase in size. However, the search for large primes, which are useful for cryptography, can be accelerated by reducing the definition set a priori (sieve methods, Mersenne numbers, etc.).

Here we study the density of primes in quadratic progressions. We will extensively use properties proved in [6]. Finally, with a reformulated and stronger version of a conjecture by Shanks [2] on the density of primes in $E_c$, we describe an algorithm to generate large primes of fixed size, and study it empirically.

## 1.2. Definitions and notations

1- If $x \in \mathbb{N}$ and $A = \{a_1 < a_2 < \cdots\} \subset \mathbb{N}$ has at least $x$ elements, then $A^{(x)} = \{a_1 \ldots a_x\}$ is the set of its first $x$ elements.

2- The density function of a set $B$ in (or conditional to) an <u>infinite</u> set $A$ is given by:
$$d_{B|A}(x) = \frac{\left|A^{(x)} \cap B\right|}{x}.$$
We will call any equivalent of $d_{B|A}$ in $+\infty$ the <u>asymptotic density</u> of $B$ in $A$.

3- The set of prime numbers will be denoted $\mathbb{P} = \{p_0 = 2, p_1 = 3, \ldots\}$. We also define the <u>primorial</u> $n\# = \prod_{\substack{p \in \mathbb{P} \\ p \leq n}} p$.

4- The (positive) divisors of an integer $x$ will be denoted by $\mathcal{D}(x)$, and by extension for any $E \subset \mathbb{Z}$ we let $\mathcal{D}(E) = \bigcup_{x \in E} \mathcal{D}(x)$. We also define the set of prime divisors: $\mathcal{D}_p(E) = \mathcal{D}(E) \cap \mathbb{P}$.

5- If $F \subset \mathbb{N}^*$ (generally a finite set of prime numbers), we denote the set of integers which are <u>coprime with every element in $F$</u> by $\mathcal{P}(F)$.

6- For any integers $(n, A) \in \mathbb{Z} \times \mathbb{N}^*$ we denote by $\{n\}_A$ the remainder of the Euclidean division of $n$ by $A$.

7- We define, for any $X \in \mathbb{Z}$, the index $j(X)$ as the quotient of its Euclidean division of $X$ by 2, i.e.:
$$X = 2j(X) + \{X\}_2$$

<u>Remark</u>: Indices were introduced in [4]. An application was studied in [5].

8- For any $a \in \mathbb{N}^*$, $b \in \mathbb{N}$, $c \in \mathbb{N}^*$ et $x \in \mathbb{N}^*$, we define the following sets:
$$L_{a,b} = \{a\xi + b, \xi \in \mathbb{N}\} = \{n \in \mathbb{N} | n \equiv b \ [a]\}$$
$$L_{a,b}^{(x)} = L_{a,b} \cap [0, ax + b)$$
$$E_c = \{\xi^2 + c, \xi \in (2\mathbb{Z} + (c-1))\}$$
$$E_c^{(x)} = E_c \cap [0, (2x+1)^2 + c)$$

## 2. Density of prime numbers in a linear and quadratic progression

### 2.1. Linear progression

The prime number theorem can be reformulated as an asymptotic density result.

**Theorem 2.1.1**: The asymptotic density of $\mathbb{P}$ is:
$$d_{\mathbb{P}|\mathbb{N}}(x) \sim \frac{1}{\ln(x)}.$$

The search for prime numbers is therefore doubly penalized, first by the fact that prime numbers are getting scarcer as they grow, and second because primality tests get more time-consuming.

It is therefore natural to look to increase this density by searching a smaller set than $\mathbb{N}$. A first approach is to look in arithmetic progressions, i.e. sets of the form $L_{a,b}$. For these sets, we have Dirichlet's arithmetic progression theorem [1], generalized by Chebotarev's theorem [3]:

**Theorem 2.1.2**: Let $\varphi$ be Euler's totient function. We have:
$$d_{\mathbb{P}|L_{a,b}}(x) \sim \frac{a}{\varphi(a)} \cdot \frac{1}{\ln(x)}$$
whenever $a, b$ are coprime.

As a result, no arithmetic progression contains only prime numbers, and these tend to get scarcer in $L_{a,b}$ at the same convergence rate as in $\mathbb{N}$. Nonetheless, it is relatively easy to find arithmetic progressions whose terms are all coprime with a finite subset $F$ of $\mathbb{P}$. To do this, we just need to solve a system of congruences of the form:
$$N \equiv a_p \ [p]$$
for any $p \in F$, with $a_p$ coprime with $p$. Solutions can be provided by the Chinese remainder theorem or by the *normalizer* method described in [10]. They correspond to $\prod_{p \in F}(p-1)$ linear progressions with the common difference $p_F \coloneqq \prod_{p \in F} p$.

Theorem 2.1.2 applies to each of these linear progressions and moreover we identify $\varphi(p_F) = \prod_{p \in F}(p-1)$. The asymptotic density of primes conditional to each of these progressions, hence also conditional on their union $\mathcal{P}(F)$, is therefore:
$$d_{\mathbb{P}|\mathcal{P}(F)} = \frac{h_F}{\ln(x)}$$
with:
$$h_F = \frac{p_F}{\varphi(p_F)} = \prod_{p \in F} \frac{p}{p-1}.$$

The asymptotic density given by the prime number theorem therefore increases by the multiplicative factor $h_F$, which can theoretically be chosen as large as desired due to the known divergence of $\sum_{p \in \mathbb{P}} \frac{1}{p}$, which implies:

$$\prod_{p \in \mathbb{P}} \frac{p}{p-1} = +\infty.$$

In $\mathbb{N}$, the asymptotic density of $\mathcal{P}(F)$ is $\frac{1}{h_F}$. Dirichlet's theorem can therefore be formulated as an independence result:

$$d_{\mathbb{P}|\mathcal{P}(F)} \sim \frac{d_{\mathbb{P}|\mathbb{N}}}{d_{\mathcal{P}(F)|\mathbb{N}}}$$

It is in this context that « basic » sieving is used to determine primes, i.e., eliminating multiples of primes as they are found. This method will be referred to as « *Sieve 1* » in this article. We also notice that primality tests are simplified in this case by the fact that by construction, elements of $F$ cannot be divisors of any element in $\mathcal{P}(F)$.

**Definition 2.1.1**:

For $N \in \mathbb{N}^*$, we define $m_N = 1 + \lfloor \log_{10} N \rfloor$ the number of decimal digits of $N$ and we call $s(N) = m_N \ln(10)$ the (natural logarithmic) <u>size of $N$</u>.

**Remark 2.1.1**: The search for prime numbers of the form $ax + b$ and of fixed size $s$ can be performed in two ways:

- Either by choosing $b$ of size $s$ and varying $x$ between 0 and $e^s/a$.
- Or by choosing $a$ and $b$ of size strictly less than $s$ and looking for $x$ of size $s - s(a)$.

In the first method, as $b$ changes with the desired size $s$, it is difficult to give an asymptotic result.

The second method consists in searching for $x \in \left[e^{s-s(a)-\ln(10)}, e^{s-s(a)}\right)$ thus, according to Dirichlet's theorem, for $s \gg 1$ the density in the search interval will be $\frac{a}{\varphi(a)} \cdot \frac{1}{s}$.

## 2.2. Quadratic progression

### 2.2.1. The case of $X^2 + c$

In this section, we are interested in the set $E_c = \{X^2 + c, X \in 2\mathbb{Z} + r\}$ where $c \in \mathbb{N}^*$ and $r = 1 - \{c\}_2$. In [6], we characterised multiples in $E_c$ of the numbers $p \in \mathcal{D}_p(E_c)$ by showing that their index $j(X) = (X - r)/2$ follows one of two arithmetic progressions:

$$\begin{cases} j(X) = j_{p,c}(0) + p.k \\ j(X) = (p - j_{p,c}(0) - r) + p.k \end{cases}$$

with the first multiple of $p$ in $E_c$ denoted by $X_{p,c}(0) \coloneqq 2j_{p,c}(0) + r$ (see definition 2.2.2a below). The two progressions above are identical if and only if $p|c$.

As in section 2.1, let $F$ be a subset of $\mathbb{P} \setminus \{2\}$ (we note that by construction, 2 never divides any element in $E_c$). The following proposition characterises the values of $c$ such that $\mathcal{D}_p(E_c)$ contains no element of $F$:

**Proposition 2.2.1a:** There exists an <u>infinite number</u> of integers $c$ of fixed parity such that $F \cap \mathcal{D}_p(E_c) = \emptyset$, which corresponding to $\prod_F \frac{p-1}{2}$ linear progressions of common difference equal to $2p_F$.

<u>Proof</u>: The desired property on $c$ is true if and only if for any $p \in F$ and $X \in \mathbb{N}$, $X^2 \not\equiv -c \ [p]$, to which we must add the parity condition.

Using Legendre symbol, this is equivalent to $\left(\frac{-c}{p}\right) = -1$ for all $p$. Now we know that there are $\frac{p-1}{2}$ quadratic nonresidues modulo $p$, so we can conclude by again using the Chinese remainder theorem.

<u>Note</u>: In section 4, we present a practical method for generating quadratic nonresidues modulo $p$.

**Remark 2.2.1a:** The search for prime numbers of the form $X^2 + c$ and of fixed size $s$ can also be performed in two ways:

- Either by choosing $c$ of size $s$ and varying $X$ between 0 and $\sqrt{c}$.
- Or by choosing $c$ of size less than $s$ and looking for $X$ of size $\frac{s}{2}$.

We will only focus on the first method.

The determination of primes in $E_c$ can be performed with a sieve method based on modular arithmetic as described in [6]. The density of residual primes in $E_c$ is bounded by that of the non-multiples of the primes already known, which can be computed using the following proposition:

**Proposition 2.2.1b:** The asymptotic density of elements in $E_c$ coprime with $F \subset \mathbb{P} \setminus \{2\}$ is given by:

$$d_{\mathcal{P}(F)|E_c}(x) \sim \prod_{p \in F}\left(1 - \frac{1}{p}\left(1 + \left(\frac{-c}{p}\right)\right)\right) = \prod_{p \in F}\left(1 - \frac{t_p}{p}\right), \text{ where } t_p = \begin{cases} 0 \text{ if } p \notin \mathcal{D}(E_c) \\ 1 \text{ if } p|c \\ 2 \text{ otherwise} \end{cases}$$

<u>Proof</u>: This is a consequence of the number of square roots of $-c$ modulo $p$ and the Chinese remainder theorem. It also yields perfect equality for multiples of $p_F$:

$$d_{\mathcal{P}(F)|E_c}(p_F x) = \prod_{p \in F}\left(1 - \frac{1}{p}\left(1 + \left(\frac{-c}{p}\right)\right)\right)$$

The sieve described in [6] will be referred to as « *Sieve 2* » henceforth.

### 2.2.2. Quadratic forms corresponding to the sieve

**Definition 2.2.2a:** For $A \in \mathcal{D}(E_c)$, let $j_{A,c}(0)$ be the index of the smallest $X \coloneqq X_{A,c}(0)$ such that $A$ divides $N(X, c) = X^2 + c$.

### a. Modulo p

**Proposition 2.2.2b:** Let $F \subset \mathcal{D}_p(E_c)$. Then $E_c \cap \mathcal{P}(F)$ is the disjoint union of $\prod_{p \in F}(p - t_p)$ distinct quadratic progressions on $x$ with parameter $b$:
$$4p_F^2 x^2 + 4bp_F x + b^2 + c, x \in \mathbb{N}$$

*Proof*: $X^2 + c$ is in $E_c \cap \mathcal{P}(F)$ if and only if $X \equiv r\,[2]$ and for any $p \in F$, $X^2 \not\equiv -c\,[p]$. Using the notations given in 2.2.1, the second condition is equivalent to $X \not\equiv \pm(2j_{p,c}(0) + r)\,[p]$.

According to the Chinese remainder theorem, there are exactly
$$\prod_{p \in F}(p - t_p)$$
solutions to these equations modulo $2p_F$. We deduce that $X^2 + c$ is in $E_c \cap \mathcal{P}(F)$ if and only if $X$ is of the form $2p_F x + b$, with $x \in \mathbb{N}$ and $b$ any solution chosen between 0 and $2p_F - 1$. We thus have, for a given choice of $b$:
$$X^2 + c = (2p_F x + b)^2 + c = 4p_F^2 x^2 + 4bp_F x + b^2 + c.$$

**Corollary 2.2.2b:** Let $F \subset \mathbb{P} \setminus \{2\}$ and $F' \subset \mathcal{D}_p(E_c)$ non-empty such that $F' \cap F = \emptyset$. For $x \to \infty$, we have:
$$d_{\mathcal{P}(F')|E_c \cap \mathcal{P}(F)}(x) \sim d_{\mathcal{P}(F')|E_c}(x) < 1$$
$$d_{\mathcal{P}(F \cup F')|E_c}(x) < d_{\mathcal{P}(F)|E_c}(x)$$

*Proof*: This is a consequence of propositions 2.2.1b and 2.2.2b. Since we have assumed that $F' \cap F = \emptyset$, it follows that:
$$d_{\mathcal{P}(F \cup F')|E_c}(x) = d_{\mathcal{P}(F)|E_c}(x) \times d_{\mathcal{P}(F')|E_c}(x)$$
from which we deduce the second inequality.

Furthermore, the Chinese theorem ensures that for $x$ a multiple of $p_F p_{F'}$, we have:
$$d_{\mathcal{P}(F')|E_c \cap \mathcal{P}(F)}(x) = d_{\mathcal{P}(F')|E_c}(x) < 1$$
which yields the first result.

**Remark 1**: Since $\mathcal{P}(F)$ contains all primes except those in $F$ (assumed finite), if $E_c \cap \mathbb{P}$ is infinite, then so is $E_c \cap \mathcal{P}(F) \cap \mathbb{P}$. We can then verify:
$$d_{\mathbb{P}|E_c \cap \mathcal{P}(F)}\left(x \prod_{p \in F}\left(1 - \frac{t_p}{p}\right)\right) = \frac{d_{\mathbb{P} \setminus F|E_c}(xp_F)}{d_{\mathcal{P}(F)|E_c}(xp_F)} = \frac{d_{\mathbb{P} \setminus F|E_c}(xp_F)}{\prod_{p \in F}\left(1 - \frac{t_p}{p}\right)}.$$

This allows us to deduce an equivalent of $d_{\mathbb{P}|E_c \cap \mathcal{P}(F)}(x)$ from an equivalent of $d_{\mathbb{P}|E_c}$ (see section 3, remark 3).

**Remark 2.2.2b:** We have $s((2p_F x + b)^2 + c) = \max\left(2(s(2p_F) + s(x)), s(c)\right)$. This size is constant equal to $s(c)$ as long as $x < \frac{\sqrt{c}}{2p_F}$.

### b. The divisors

Let $A_1$ be an element of $\mathcal{D}(E_c)$. Then, for $X$ such that $A_1 | X^2 + c$:
$$X^2 + c = A_1 B.$$

The set of integers $B$ is also a subset of $\mathcal{D}(E_c)$. A subset of such integers $B$ is given by the following two quadratic progressions.

**Proposition 2.2.2b1:** Let $B_{A_1,c} := \frac{1}{A_1}(X_{A_1,c}(0)^2 + c)$. Then $\mathcal{B}_{A_1,c} := \left(\frac{1}{A_1}E_c\right) \cap \mathbb{N}$ contains the progressions:

$$\mathcal{B}_{A_1,c,\varepsilon} = \left\{4n\left(A_1 n + \varepsilon X_{A_1,c}(0)\right) + B_{A_1,c}, n \in \mathbb{N}\right\}$$

for $\varepsilon \in \{\pm 1\}$.

When $A_1 \in E_c$, we have $B_{A_1,c} = 1$.

The quadratic progression $\mathcal{B}_{A_1,c,\varepsilon}$ corresponds to an irreducible element of $\mathbb{Z}[n]$ if and only if $\gcd(A_1, X_{A_1,c}(0), B_{A_1,c}) = 1$.

*Proof*: In [6], we showed that for each divisor $A_1 \in \mathcal{D}(E_c)$, there are at least two arithmetic progressions of indices generating multiples of $A_1$ given by:
$$j_{A_1,c,A_1}(n) = A_1 n + j_{A_1,c}(0) \text{ et } j_{A_1,c,1}(n) = A_1 n + j_{A_1,c,1}(0).$$

We also have $j_{A_1,c,1}(0) + j_{A_1,c}(0) \equiv -r\ [A_1]$, i.e. $X_{A_1,c,1}(0) = 2j_{A_1,c,1}(0) + r \equiv -X_{A_1,c}(0)\ [2A_1]$. By substituting $X = 2j + r$ in $X^2 + c$, we get the two quadratic progressions (with the same first term):

$$X^2 + c = A_1\left(4n\left(A_1 n \pm X_{A_1,c}(0)\right) + B_{A_1,c}\right).$$

Furthermore, if $A_1 \in E_c$ it is clear that $A_1 = X_{A_1,c}(0)^2 + c$ hence $B_{A_1,c} = 1$.

Finally, as the polynomial $X^2 + c$ has no real root, the same applies to $n \mapsto 4n\left(A_1 n \pm X_{A_1,c}(0)\right) + B_{A_1,c}$. It therefore corresponds to an irreducible element of $\mathbb{Z}[n]$ if and only if its coefficients are coprime, and since $B_{A_1,c}$ is odd, this is equivalent to:

$$\gcd(A_1, X_{A_1,c}(0), B_{A_1,c}) = 1.$$

**Proposition 2.2.2b2:** Let $A \in \mathcal{D}(\mathcal{B}_{A_1,c,\varepsilon})$ and $n_0$ the smallest integer such that:
$$A | 4n_0\left(A_1 n_0 + \varepsilon X_{A_1,c}(0)\right) + B_{A_1,c}.$$

For any element $AB = 4n\left(A_1 n + \varepsilon X_{A_1,c}(0)\right) + B_{A_1,c}$ of $\mathcal{B}_{A_1,c,\varepsilon} \cap A\mathbb{Z}$, there exists a divisor $a$ of $A$ and integers $u, n_{A,a,\varepsilon}(0)$ and $k$ such that:

$$\gcd\left(A_1, \frac{A}{a}\right) | A_1 n_0 + \varepsilon X_{A_1,c}(0)$$

$$A_1 u \equiv \gcd\left(A_1, \frac{A}{a}\right)\left[\frac{A}{a}\right]$$

$$\gcd\left(A_1 a, \frac{A}{a}\right) \Big| \left(A_1 u + \gcd\left(A_1, \frac{A}{a}\right)\right) n_0 + \varepsilon X_{A_1,c}(0)$$

$$n = n_{A,a,\varepsilon}(0) + k \cdot \frac{A}{\gcd\left(aA_1, \frac{A}{a}\right)}$$

*Proof*: Assume that $A$ divides $4n\left(A_1 n + \varepsilon X_{A_1,c}(0)\right) + B_{A_1,c}$. We can then write:

$$A | 4n\left(A_1 n + \varepsilon X_{A_1,c}(0)\right) - 4n_0\left(A_1 n_0 + \varepsilon X_{A_1,c}(0)\right) = 4(n - n_0)\left(A_1(n + n_0) + \varepsilon X_{A_1,c}(0)\right)$$

Thus there is a factorization $A = ab$ such that $a | n - n_0$ and $b | A_1(n + n_0) + \varepsilon X_{A_1,c}(0)$, i.e. $n \equiv n_0 \, [a]$ and $A_1 n \equiv -\left(A_1 n_0 + \varepsilon X_{A_1,c}(0)\right) \, [b]$.

The second equation has solutions only if $\gcd(A_1, b) \,|\, A_1 n_0 + \varepsilon X_{A_1,c}(0)$. Under this condition, let $u$ be such that $\frac{A_1}{\gcd(A_1,b)} u \equiv 1 \left[\frac{b}{\gcd(A_1,b)}\right]$. The second equation is equivalent to:

$$n \equiv -u \frac{A_1 n_0 + \varepsilon X_{A_1,c}(0)}{\gcd(A_1, b)} \left[\frac{b}{\gcd(A_1, b)}\right].$$

The system then has solutions if and only if:

$$n_0 \equiv -\frac{u}{\gcd(A_1, b)}\left(A_1 n_0 + \varepsilon X_{A_1,c}(0)\right) \left[\gcd\left(a, \frac{b}{\gcd(A_1, b)}\right)\right].$$

This can be rewritten as follows:

$$\gcd(\gcd(aA_1, ab), b) = \gcd(aA_1, b) \,|\, (A_1 u + \gcd(A_1, b)) n_0 + \varepsilon X_{A_1,c}(0).$$

Under these conditions, the remainder of $n$ modulo $\frac{A}{\gcd\left(aA_1, \frac{A}{a}\right)}$ is uniquely determined. Let us denote it by $n_{A,a,\varepsilon}(0)$, then we must have $n = n_{A,a,\varepsilon}(0) + k \cdot \frac{A}{\gcd\left(aA_1, \frac{A}{a}\right)}$ for some $k \in \mathbb{N}$.

**Definition 2.2.2b2**: Subject to the conditions of the previous proposition, we denote by $\left(n_{A,a,\varepsilon}(k)\right)$ the above arithmetic progression, i.e.: $n_{A,a,\varepsilon}(k) := n_{A,a,\varepsilon}(0) + k \cdot \frac{A}{\gcd\left(a, \frac{A}{a}\right)}$.

**Proposition 2.2.2b3**: The two progressions $\left(n_{A,A,1}(k)\right)$ and $\left(n_{A,A,-1}(k)\right)$ always exist and have the same common difference $A$: we will call them dual progressions.

*Proof*: The existence conditions of $\left(n_{A,a,\varepsilon}(k)\right)$ are as follows.

$$\gcd\left(A_1, \frac{A}{a}\right) \,|\, A_1 n_0 + \varepsilon X_{A_1,c}(0)$$

$$A_1 u \equiv \gcd\left(A_1, \frac{A}{a}\right) \left[\frac{A}{a}\right]$$

$$\gcd\left(A_1 a, \frac{A}{a}\right) \,|\, \left(A_1 u + \gcd\left(A_1, \frac{A}{a}\right)\right) n_0 + \varepsilon X_{A_1,c}(0)$$

For $a = A$, all these conditions are trivially verified, and the common difference of the progression obtained is $\frac{A}{\gcd(AA_1, 1)} = A$.

**Remark 2.2.2b1**: We have $\mathcal{D}_p(\mathcal{B}_{A_1,c,\varepsilon}) \subseteq \mathcal{D}_p(E_c)$ and for any prime divisor $p$ that is not a divisor of $A_1$, we have as many dual progressions of multiples of $p$ in the set $\mathcal{B}_{A_1,c,\varepsilon}$ as in the set $E_c$, i.e. $t_p$ progressions in both sets.

*Proof*: If $p$ prime, there are $t_p$ dual sequences in $E_c$. But if $p$ is prime with $A_1$, since $\mathcal{B}_{A_1,c,\varepsilon}$ corresponds to an arithmetic progression with the common difference $A_1$ on $X$, by the Chinese remainder theorem, it must have the same number of arithmetic sub-progressions with the common difference $p$ (prime with $A_1$) corresponding to multiples of $p$.

**Corollary 2.2.2b3**: If $A_1$ and $A$ are coprime, any progression $\left(j_{A,c,a}(n')\right)$ of indices of $E_c$ corresponds to a sequence of $\mathcal{B}_{A_1,c,\varepsilon}$ of the same constant asymptotic density.

*Proof*: Since $A_1$ is coprime with $A$, it is also coprime with the common difference of the arithmetic progression $j_{A,c,a}(n')$, and as stated in proposition 2.2.2b1, $\mathcal{B}_{A_1,c,\varepsilon}$ corresponds to an arithmetic progression of indices with common difference $A_1$. The Chinese theorem yields that $\mathcal{B}_{A_1,c,\varepsilon}$ thus contains terms corresponding to $\left(j_{A,c,a}(n')\right)$ occurring at the same frequency, i.e. with the same asymptotic density equal to the inverse of the common difference of $\left(j_{A,c,a}(n')\right)$ i.e. $\gcd\left(a, \frac{A}{a}\right)/A$.

**Proposition 2.2.2b5**: Let $p \in \mathcal{D}_p(E_c)$ and $\nu$ be the maximal power such that $p^\nu | c$. With $A_1 = p^\nu$, we have $\mathcal{B}_{A_1,c,1} = \mathcal{B}_{A_1,c,-1}$ and if $p \in \mathcal{D}_p(\mathcal{B}_{A_1,c,1})$ then $\nu$ is even and we obtain:

$$\mathcal{D}_p(\mathcal{B}_{A_1,c,1}) = \mathcal{D}_p(E_c),$$

otherwise

$$\mathcal{D}_p(\mathcal{B}_{A_1,c,1}) = \mathcal{D}_p(E_c) \setminus \{p\}.$$

*Proof*: This result can be directly deduced from proposition 2.2.2 of [6] and corollary 2.2.2b3.

**Corollary 2.2.2b5**: We have $\mathcal{D}_p(\mathcal{B}_{A_1,c,\varepsilon}) \subseteq \mathcal{D}_p(E_c)$ and $\lim\limits_{x \to \infty} d_{\mathcal{P}(F)|\mathcal{B}_{A_1,c,\varepsilon}}(x) \geq \lim\limits_{x \to \infty} d_{\mathcal{P}(F)|E_c}(x)$.

*Proof*: Since $\mathcal{B}_{A_1,c,\varepsilon}$ consists of divisors of elements of $E_c$, we clearly have $\mathcal{D}_p(\mathcal{B}_{A_1,c,\varepsilon}) \subseteq \mathcal{D}_p(E_c)$. Corollary 2.2.2b3 states that, if $F \cap \mathcal{D}_p(A_1) = \emptyset$, $\lim\limits_{x \to \infty} d_{\mathcal{P}(F)|\mathcal{B}_{A_1,c,\varepsilon}}(x) = \lim\limits_{x \to \infty} d_{\mathcal{P}(F)|E_c}(x)$. But, even if $F \cap \mathcal{D}_p(A_1) \neq \emptyset$ the Chinese theorem leads to $\lim\limits_{x \to \infty} d_{\mathcal{P}(F)|\mathcal{B}_{A_1,c,\varepsilon}}(x) \geq \lim\limits_{x \to \infty} d_{\mathcal{P}(F)|E_c}(x)$.

Similarly to <u>section 2.1</u>, let $F \subset \mathbb{P} \setminus \{2\}$ be a finite set and let $F' \subset \mathcal{D}_p(\mathcal{B}_{A_1,c,\varepsilon})$ be another non-empty set such that $F' \cap F = \emptyset$.

**Proposition 2.2.2b6**: There is a finite number of sub-progressions of $\mathcal{B}_{A_1,c,\varepsilon}$ corresponding to arithmetic progressions with common difference $p_{F'}$ on the index, the terms of which have no prime divisors in $F'$. If $F'$ is also disjoint from $\mathcal{D}_p(\{A_1\})$, there are exactly $\prod_{F'}(p - t_p)$ sub-progressions and we have:

$$\lim_{x \to \infty} d_{\mathcal{P}(F \cup F')|\mathcal{B}_{A_1,c,\varepsilon}}(x) < \lim_{x \to \infty} d_{\mathcal{P}(F)|\mathcal{B}_{A_1,c,\varepsilon}}(x)$$

<u>Proof</u>: The terms of $\mathcal{B}_{A_1,c,\varepsilon}$ that are multiples of $p \in F'$ are given by an equation of degree 2 on the index, which therefore has at most two solutions. If we assume additionally that $p$ does not divide $A_1$, corollary 2.2.2b3 yields that the number of solutions is exactly $t_p$. We conclude as usual with the Chinese remainder theorem, and the strict inequality comes simply from the observation $\prod_{F'}(p - t_p) < p_{F'}$.

**Remark 2**: It is possible to generalize <u>remark 1</u> to this case.

In [7] it was shown that there are infinitely many pairs of primes $(p, q) \in \mathcal{D}_p(E_c) \times \mathcal{D}_p(E_c)$ such that $N(X, c) = pq \in E_c$. Below we present a hypothesis on the infinity of pairs of primes $(p, q)$ with $p$ fixed.

**Conjecture 1**: Let $p \in \mathcal{D}_p(E_c)$ such that $p$ not divide $c$. There are infinitely many $q \in \mathcal{D}_p(E_c)$ such that $pq \in E_c$.

Equivalently, $\mathcal{B}_{p,c,\varepsilon}$ contains infinitely many primes for at least one value of $\varepsilon$.

Below we present some empirical results for $c = 1$. We take $p \coloneqq 5 \in E_1$. We then obtain $X_{5,1}(0) = 2$ and $B_{5,1} = 1$. We count the number of primes in $\mathcal{B}_{5,1,\varepsilon}$ corresponding to elements of $E_c^{(x)}$, i.e. the prime elements of $B(n_\varepsilon)$ such that $n_\varepsilon \leq \left\lfloor \frac{x}{5} \right\rfloor$. In graph 1 below, for $\varepsilon = 1$, we plot $\left|\mathbb{P}^{(x)} \cap \mathcal{B}_{5,1,1}\right| = x. d_{\mathcal{B}_{5,1,1}|\mathbb{P}}(x)$ as a function of $x$. Details of the Maple options used to obtain the relationships presented in this section are given in Appendix A.

**graph 1:** $\left|\mathbb{P}^{(x)} \cap \mathcal{B}_{5,1,1}\right|$ as a function of $x$

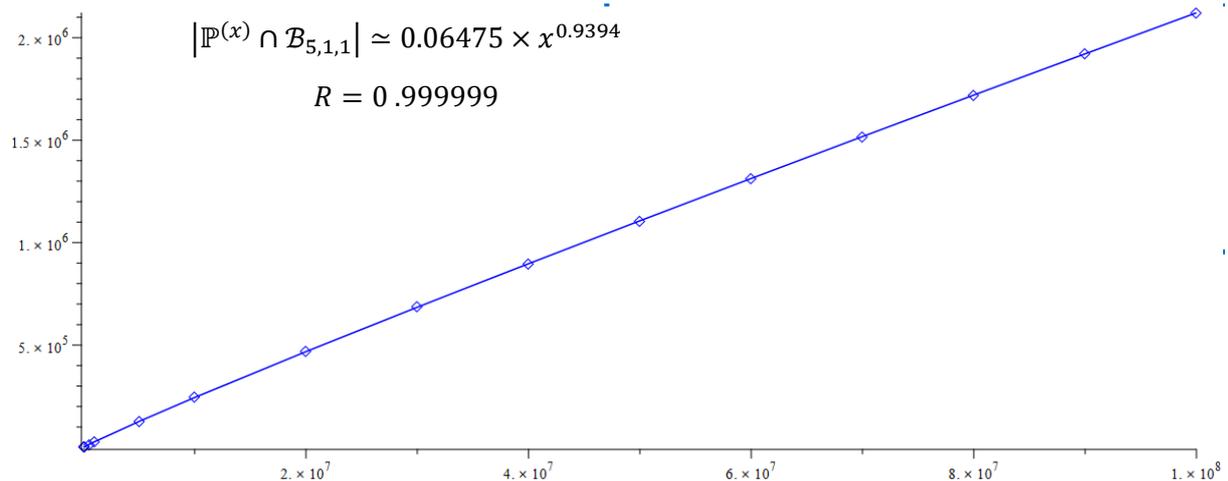

The evolution of the number of elements of $\mathcal{B}_{5,1,\varepsilon}$ in $\mathbb{P}^{(x)}$ is of the same form as those of $E_1$, i.e. $ax^b, b > 0$, which we studied in [6] and subsequently corroborated Landau's conjecture (namely the infinity of $\mathbb{P} \cap E_1$) and more generally Shanks' conjecture. Here we extend this conjecture; that is, we conjecture that $\mathcal{B}_{5,1,1} \cap \mathbb{P}$ is infinite. Similar results were obtained for $\varepsilon = -1$ as well as for all divisors and all values of $c \geq 1$ studied.

## 3. Shanks' conjecture [2]

The hypothesis that there are infinitely many primes in a quadratic form was first emitted by Landau in 1912, for $X^2 + 1$. In 1923, Hardy and Littlewood proposed two conjectures in [8], E and F, for the forms $X^2 + 1$ and $aX^2 + bX + c$ respectively. In 1960, Shanks rewrote the F conjecture for the form $X^2 + c$ in [2]. To date, none of these conjectures has been proved.

In this section, we propose to rewrite Shanks' conjecture [2] with the properties of this article and [6], and more precisely with "*Sieve* 1" and "*Sieve* 2".

**Conjecture 2**: *For any integer $c \in \mathbb{N}^*$, we have the following asymptotic density:*

$$d_{\mathbb{P}|E_c}(x) \sim h_c \frac{1}{\ln(x)}$$

where $h_c = \prod_{p \in \mathbb{P}\setminus\{2\}} \frac{p-1-\left(\frac{-c}{p}\right)}{p-1} = \prod_{p \in \mathbb{P}\setminus\{2\}} \frac{p-t_p}{p-1}$ with $t_p = \begin{cases} 0 \text{ si } p \notin \mathcal{D}_p(E_c) \\ 1 \text{ si } p \in \mathcal{D}(c) \\ 2 \text{ sinon.} \end{cases}$

**Explanation**: The proof that the infinite product defining $h_c$ converges is given in [9]. When $p \in \mathcal{D}(c)$, we have $t_p = 1$ which gives a factor equal to 1, so we can exclude these values from the calculation of $h_c$. The equality

$$t_p = \left(\frac{-c}{p}\right) + 1$$

is immediate since the prime divisors of $E_c$ are the $p$ such that $-c$ is a quadratic residue modulo $p$.

We estimated $h_c$ for $c \in \{1,3,5\}$ by truncating the product to primes less than $4.10^6$, using the algorithm from [6] to compute $t_p$. The estimated values of $h_c$ for $c \in \{1,3\}$ are close to those supplied by Shanks [2]:

| $c$ | $h_c$ from [2] | $h_c$ estimated | Difference |
|---|---|---|---|
| 1 | 1.372813 | 1.372771 | 0.000042 |
| 3 | 1.120733 | 1.120727 | 0.000006 |
| 5 | - | 0.528219 | - |

We know from <u>proposition 2.2.1b</u> that for any finite set $F \subset \mathbb{P}$, we have the following asymptotic density:

$$d_{\mathcal{P}(F)|E_c} = \prod_{p \in F\setminus\{2\}} \left(1 - \frac{t_p}{p}\right)$$

We also know that for any $b$ prime with $p_F$, we have:

$$d_{\mathbb{P}|L_{p_F,b}}(x) \sim \frac{p_F}{\varphi(p_F).\ln(x)}$$

Moreover, $\mathcal{P}(F)$ consists of a finite number of arithmetic progressions with common difference $p_F$ on $x$, so we also have:

$$d_{\mathbb{P}|\mathcal{P}(F)}(x) \sim \frac{p_F}{\varphi(p_F).\ln(x)} = \left(\prod_{p\in F} \frac{p}{p-1}\right)\frac{1}{\ln(x)}$$

By conjecturing an "independence" between $E_c$ and $\mathcal{P}(F)$ when $F$ tends towards $\mathbb{P}$, we obtain the following result:

$$d_{\mathbb{P}|E_c}(x) \sim \left(\prod_{p\in\mathbb{P}\setminus\{2\}} \left(1-\frac{t_p}{p}\right)\right)\left(\prod_{p\in\mathbb{P}} \frac{p}{p-1}\right)\frac{1}{\ln(x^2)} = h_c \frac{1}{\ln(x)}$$

This conjecture combines thus both *"Sieve 1"* (or Dirichlet's arithmetic progression theorem) and *"Sieve 2"*.

According to this conjecture, it would be interesting *a priori* to look for values of $c$ such that $h_c$ is large to increase the asymptotic density of primes in $E_c$.

**Remark 3**: A generalized *"independence"* conjecture would also make it possible to extend Shanks' conjecture to the subsets of $E_c$ studied in Part 2, for instance to calculate $d_{\mathbb{P}|E_c \cap \mathcal{P}(F)}(x)$, or as in the following hypothesis:

**Hypothesis 1**: The Shanks hypothesis applies to sets $\mathcal{B}_{A_1,c,\varepsilon}$ when it corresponds to an irreducible quadratic progression.

For the rest of the article, we will use a stronger version of Shanks' conjecture:

**Hypothesis 2**: We have:

$$\sup_c \left|\ln(X)\, d_{\mathbb{P}|E_c}(X) - h_c\right| \to_{X\to+\infty} 0.$$

**Remark 4**: Determining the number of primes of the form $X^2 + c$ for $X \in [\![0, \sqrt{c}]\!]$ will thus provide an approximation of $h_c$ when $c$ is large, uniformly in $c$. Furthermore, in proposition 2.2.1a, we showed that integers $c$ such that $F \cap \mathcal{D}_p(E_c) = \emptyset$ are given by progression $c(n) = 2p_F n + c(0)$. With $n \geq 1$ fixed, the size of $c(n)$ is equal to $s(2p_F n)$ which allows us to define the same size for a finite number, equal to $\prod_F \frac{p-1}{2}$, of values of $c$.

With hypothesis 2, we can thus obtain a prime density arbitrarily close to $h_c/m_c$, over an interval of values of $X$ as large as we wish.

## 4. Solving systems of congruences on c

In this section, we present properties around solving a system of linear congruences on c ensuring that $F \cap \mathcal{D}_p(E_c) = \emptyset$,, where $F$ is a finite set of primes. We will link values of $c$ of opposite parity. These properties will be used in the algorithm determining the set of even and odd values of $c$.

**Definition 4.1**: Let $p$ be an odd prime number.

We define the sets of quadratic residues and non-residues modulo $p$, respectively $E_p^{RQ}$ and $E_p^{NRQ}$ which form a partition of $\mathbb{Z}/p\mathbb{Z}$:

$$E_p^{RQ} = \{b \in \mathbb{Z}/p\mathbb{Z} | \exists x \in \mathbb{Z}, b \equiv -x^2\}$$
$$E_p^{NRQ} = \left(\frac{\mathbb{Z}}{p\mathbb{Z}}\right) \setminus E_p^{RQ}.$$

with $|E_p^{RQ}| = \frac{p+1}{2}$ and $|E_p^{NRQ}| = \frac{p-1}{2}$.

We recall that according to the proposition 2.2.1a, there exist <u>infinitely</u> many integers $c$, such that $F \cap \mathcal{D}_p(E_c) = \emptyset$. This is exactly the set $\bigcap_{p \in F} E_p^{NRQ}$. We denote $\mathcal{C}_F$ the set of remainders of $\bigcap_{p \in F} E_p^{NRQ}$ modulo $2p_F$, and $\mathcal{C}_{F,r}$ this set with the additional condition $\{c\}_2 = 1 - r$ with $r \in \{0,1\}$. $\mathcal{C}_{F,0}$ and $\mathcal{C}_{F,1}$ have the same cardinal $T_F = \prod_{p \in F} \frac{p-1}{2}$.

We denote $\mathcal{C}_{F,r} = \{c_{F,r,1} \dots c_{F,r,T_F}\}$.

**Proposition 4.1**: $\mathcal{C}_{F,0}$ and $\mathcal{C}_{F,1}$ are equal modulo $p_F$.

*Proof*: This is another consequence of the Chinese remainder theorem. Since $p_F$ is odd, each congruence class modulo $p_F$ has exactly one even and one odd representative between 0 and $2p_F - 1$. For any $i \in [\![1, T_F]\!]$ there exists a unique $i' \in [\![1, T_F]\!]$ such that $c_{F,1,i} = c_{F,0,i'} + p_F[2p_F]$.

We now try to impose the smallest prime divisor $q_1 = p_k$ of $E_c$. To do this, let $F = \{p_1 \dots p_{k-1}\}$ and we look for $c$ in $\mathcal{C}_{F,r} \cap E_{p_k}^{RQ}$. We note that $j_{q_1,c}(0)$ depends only on the class of $c$ modulo $q_1$, and can therefore be determined a priori.

Similarly, we denote by $\tilde{\mathcal{C}}_{q_1}$ the remainders of $\left(\bigcap_{p \in F} E_p^{NRQ}\right) \cap E_{q_1}^{RQ}$ modulo $q_1\# = 2p_F q_1$ and $\tilde{\mathcal{C}}_{q_1,r}$ those of parity $1 - r$. $\tilde{\mathcal{C}}_{q_1,r}$ has $\frac{p_k+1}{2} \prod_{l=1}^{k-1} \frac{p_l-1}{2}$ elements.

**Remark 4.1**: Assume $c$ is odd. $N(X',c) \equiv N(X'', c + q_1) [q_1]$ when:

$$j(X') + j(X'') \equiv \frac{q_1 - 1}{2} [q_1]$$

*Proof*: We have $N(X',c) = 4j(X')^2 + c$. If we have $j(X'') \equiv \frac{q_1-1}{2} - j(X') [q_1]$, therefore:

$$N(X'', c + q_1) = (2j(X'') + 1)^2 + (c + q_1) \equiv (q_1 - 2j(X'))^2 + c [q_1]$$

which yields the result.

We now explain how to solve the Diophantine equations that determine $\mathcal{C}_F$ and $\tilde{\mathcal{C}}_{q_1}$ with limited use of the modulo operator.

We focus on the case of $\mathcal{C}_F$, as the extension to $\tilde{\mathcal{C}}_{q_1}$ does not present any difficulty. We write $F = \{q_1 \dots q_M\}$. By recurrence, we reduce the problem to the successive solution of the system, for $m$ going from 0 to $M - 1$:

$$(S_m) \quad \begin{cases} (i) \ c \equiv a \ [2p_{F(m)}] \\ (ii) \ c \equiv b \ [p_{m+1}] \end{cases}$$

for all values of $b \in E^{NRQ}_{q_{m+1}}$ and $a \in \mathcal{C}_{F(m)}$.

We fix $a$, $b$ and we set $d = a - b$. If we have $(x, y)$ such that:

$$(1.3) \quad p_{m+1} x - 2 p_{F(m)} y = d$$

the solution is then:

$$c = \left(2 p_{F(m)} y + a\right) \left[2 p_{F(m+1)}\right]$$

If $(u, v)$ is a solution for $d = 1$ (Bézout coefficients), then we have:

$$y = vd \; [p_{m+1}]$$

This leads to the proposition below, which gives the set of values $\mathcal{C}_F$.

**Definition 4.2**: For $m$ fixed, we note $\mathcal{C}_{F(m), r} = \{c_{m,r,1} \ldots c_{m,r,N_m}\}$.

**Proposition 4.2**: Let $E^{NRQ}_{q_{m+1}} = \{b_0 \ldots\}$. For any fixed value $c_{F(m), r, i_0}$, we get the $c_{F(m+1), r, n(i_0, j)}$ solutions of $(i)$ $c \equiv c_{F(m), r, i_0} \left[2 p_{F(m)}\right]$ and $(ii)$ $c \equiv b_j \; [q_{m+1}]$ from the value of $v$ and the solution $c_{F(m+1), r, n(i_0, 0)}$ for $b = b_0$:

$$(3.3) \quad c_{F(m+1), r, n(i_0, j)} = c_{F(m+1), r, n(i_0, 0)} + 2 p_{F(m)} y (b_0 - b_j) \left[2 p_{F(m+1)}\right]$$

**Remark 4.2**: To obtain the elements of $\tilde{\mathcal{C}}_{q_1}$, with $q_1 = p_M$, we would use $F = \{p_1 \ldots p_M\}$ and just replace $E^{NRQ}_{q_1}$ by $E^{RQ}_{q_1}$ in the last step $m = M$.

Furthermore, the pre-calculation of $y(b_0 - b_j)$ modulo $q_{m+1}$ limits the use of the modulo operator in the algorithmic implementation of (3.3).

## 5. Algorithms for calculating c with fixed smallest prime divisor

### 5.1. Presentation of the algorithms

We present a first algorithm that maps $q_1 \in \mathbb{P}$ to values of $c$ such that $q_1 = \min \mathcal{D}_p(E_c)$. The index $j_{q_1, c}(0) = \min\limits_{q_1 | (2j+r)^2 + c} j$ is also returned. The results of the previous parts allow us to crucially reduce the number of computations to determine the set of even and odd solutions $c$. In this algorithm, we first compute $\mathcal{C}_{F(m), 0}$ with the proposition 4.2, for $m = 1 \ldots M$. Then we compute the pairs $\left(c, j_{q_1, c}(0)\right)$ for $c \in \tilde{\mathcal{C}}_{q_1, 0}$. The pairs $\left(c, j_{q_1, c}(0)\right)$ for $c \in \tilde{\mathcal{C}}_{q_1, 1}$ are then obtained by the proposition 4.1 and remark 4.1.

**ALGORITHM 1** [EL1]:

A general description of the algorithm is given here:
- ➢ The inputs to the algorithm are $q_1$ and the number of values of $c$ to be returned.
- ➢ The first step initializes the variables and arrays. Then, for each prime $p_m \in \{p_1, \ldots, p_M = q_1\}$, the second and third steps are performed.

> The second step calculates the quadratic residues and nonresidues of $p_m$. Then, $v$ is then calculated by solving (1.3) for $d = 1$. The values of $y(b_0 - b_j)$ modulo $p_m$ are then precalculated (Remark 4.2).
> The third step calculates the elements of $\mathcal{C}_{F(m),0}$ by recurrence on $m = 1 \ldots M - 1$, and finally $\tilde{\mathcal{C}}_{q_1,0}$.
> The last step calculates all the pairs $\left(c, j_{q_1,c}(0)\right)$ to be returned, with $c \in \tilde{\mathcal{C}}_{q_1}$.

Algorithm 1 is exhaustive and thus performs poorly when $q_1$ is large. The search for one or a few pairs for $q_1 = 115\ 140\ 317$ requires a faster algorithm, we thus need to relax the exhaustivity target. For this value of $q_1$, we have $M = 6\ 580\ 238$ primes to process and the number of digits in c is comparable to that of the primorial $q_1\#$, i.e. $m_c$ is of the order of 50 000 006. We propose a second algorithm that computes just one quadratic nonresidue for each value of $p < p_m$, using a seed $X$.

**ALGORITHM 2** [EL2]:

A general description of the algorithm is given here:
> The inputs to the algorithm are $q_1$, the number of values of $c$ to be returned, that has to be not greater than $\frac{q_1+1}{2}$, and the seed $X$ for calculating the quadratic residue.
> The first step initializes the variables and tables.
> The second step can be run in parallel. For each prime number $p_m$, the primorial $p_m\#$ is calculated, followed by $v$ (Remark 4.1). Then, we search for a quadratic nonresidue $n_m < p_m$ such that $\left(\frac{n_m}{p_m}\right) = -1$ and we retain $b = -4X^2 n_m \bmod p_m$ for the second equation in $(S_m)$. When $p_m \equiv 3\ [4]$ we choose $n_m = -1$, when $p_m \equiv 5\ [8]$ we choose $n_m = 2$ otherwise we test the primes $3 \leq p < p_m$ and choose the first quadratic nonresidue.
> The third step calculates the quadratic residues of $q_1$ and performs the pre-calculations described in Remark 4.2.
> The last step computes all pairs $\left(c, j_{q_1,c}(0)\right)$ to be returned, with $c \in \tilde{\mathcal{C}}_{q_1}$.

## 5.2. Results

In this section, we fix $q_1 \in \mathbb{P}$ and generate some odd values of $c$ such that $q_1 = \min \mathcal{D}_p(E_c)$. The primes of $E_c$ are of the form $N(X, c) = c + X^2, X \in 2\mathbb{N}$. To obtain a prime of the same size as $c$, we impose $X \leq \sqrt{c}$.

Remark: The pairs $\left(c, j_{q_1,c}(0)\right)$ were obtained using the [EL1] or [EL2] algorithm, implemented in C# and compiled with Microsoft Visual Studio 2019. The number of primes in finite subsets of $E_c$ was counted using the *isprime* function in Maple 2022. Each value of $q_1$ corresponds to an infinite number of pairs $\left(c, j_{q_1,c}(0)\right)$. The pairs with which the results presented in this article were obtained are available in [EL3] and [EL7]. The numerical series corresponding to the graphs are available in [EL3] and [EL4].

We approximate the value of $h_c$ by $h_c(X) = m_c \ln(10)\, d_{\mathbb{P}|E_c}(X)$, which is valid if one admits hypothesis 2 (uniform version of Shanks' conjecture). The asymptotic density of primes in $E_c$

is denoted by $d_c(X) = d_{\mathbb{P}|E_c}(X)$. For $X_{max}$ the strongest value of $X$ used in a set of results, we let simply $h_c = h_c(X_{max})$ and $d_c = d_c(X_{max})$.

We will first present the evolution of the value of $h_c$ and $d_c$ as a function of the three parameters studied:
- the value of $q_1$,
- the value of $j_{q_1,c}(0)$,
- the size of $c$.

We will then study the distribution of primes in $E_c$ in an interval corresponding to $X \in [\![0, 4m_c]\!]$. Finally, we will show the interest of this method in cryptography for determining large primes.

### 5.2.1. Logarithmic evolution of h_c as a function of q₁

Remark 3 in section 3 allows us to estimate the asymptotic density $d_c$ and the value of $h_c$ from the number of primes of $E_c$ over the interval corresponding to $X \in [\![0, \sqrt{c}]\!]$. For practical reasons, we restrict ourselves to the interval $X \in [\![0, 4.10^7]\!]$.

Using Algorithm 1, we have calculated values of $c$ for values of $q_1$ such that $18 \leq m_{p_1\#} \leq 2001$, yielding $\sqrt{c} > 4.10^7$. We will first present the calculation of the estimated values of $h_c$ and $d_c$ as well as the convergence of $h_c$. We will then show the evolution of $h_c$ and $d_c$ as a function of $q_1$, which will allow us to estimate the number of primes in a fixed interval regardless of $q_1$. Next, we show that $h_c$ depends only weakly on $j_{q_1,c}(0)$ and the size of $c$. The asymptotic density $d_c$ also depends only weakly on $j_{q_1,c}(0)$, but decreases with the size of $c$, as per conjecture 2.

In the interval $[\![0, 4.10^7]\!]$, we count the primes in increments of $X$ equal to $10^4$ and cumulatively up to $X_{max} = 4.10^7$. Graph 2 shows the values of $h_c(X)$ and the estimated value $h_c \simeq 5.9860$ obtained for $q_1 = 727$ with $m_c = 301$ and $j_{q_1,c_1}(0) = 0$.

**Graph 2**: $h_c(X)$ for $q_1 = 727$ ($m_c = 301$) [EL4]

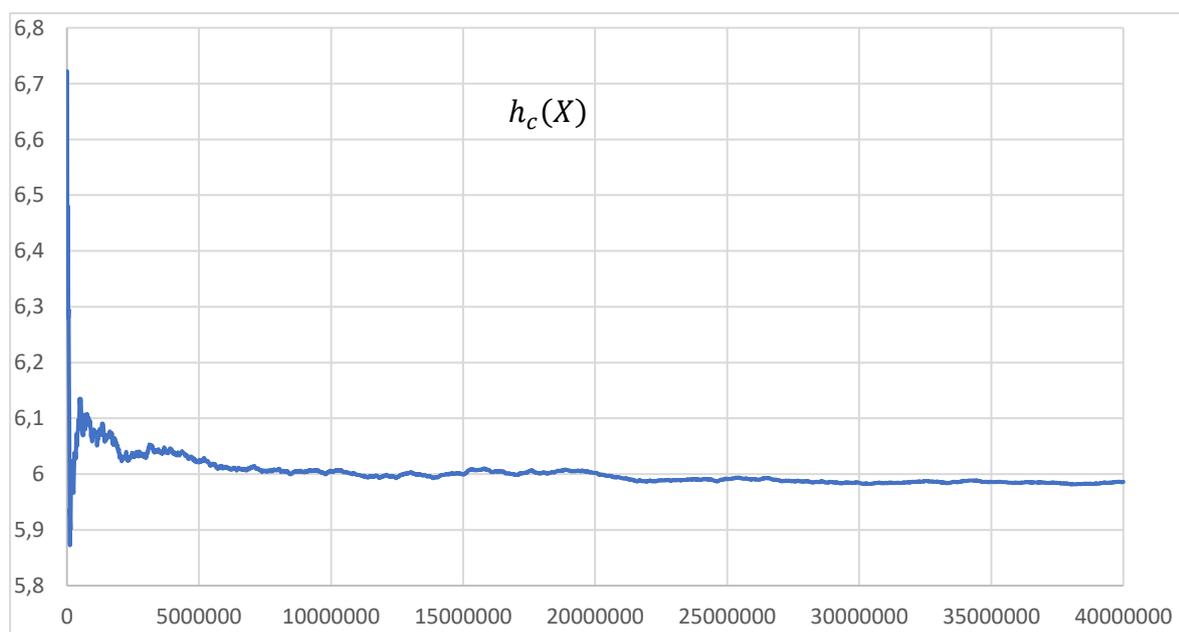

The density curve $d_c(X)$ also looks almost constant, we measure $d_c = 0,8636\ldots\%$. This corroborates the conjecture that the density $d_c$ is asymptotically related to $h_c$ via $d_c = h_c/s(c)$. This observation was confirmed for all pairs studied.

Graph 3 presents the evolution of $h_c$ and $d_c$ when $q_1$ varies. More precisely, we used the first value of $c$ such that $j_{q_1,c}(0) = 0$ given by Algorithm 1 to calculate $h_c$ and $d_c$. The numerical results are given in [EL3].

**Graph 3**: $h_c$ and $d_c$ (%) as a function of $q_1$.

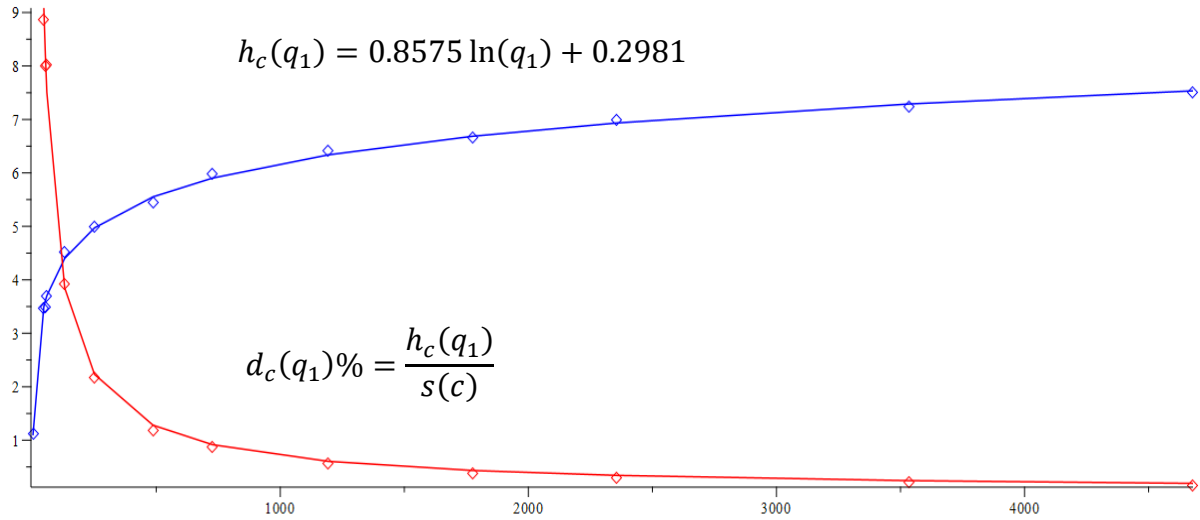

$$h_c(q_1) = 0.8575 \ln(q_1) + 0.2981$$

$$d_c(q_1)\% = \frac{h_c(q_1)}{s(c)}$$

In table 1, the regressions $h_c(q_1)$ and $d_c(q_1)$ are given for both limit values of $j_{q_1,c}(0) \in \{0, (q_1-1)/2\}$ with the correlation coefficient. The value of $h_c$ increases logarithmically with the value of $q_1$ while the density $d_c$ decreases.

**Table 1**: Regressions of $h_c(q_1)$ and $d_c(q_1)$ on $j_{q_1,c}(0)$ [EL3].

| $j_{q_1,c}(0)$ | $h_c$ and $d_c$ as function of $q_1$ | |
|---|---|---|
| 0 | (**H1**)  $h_c(q_1) = 0.8575 \ln(q_1) + 0.2981$ | R = 0,9991 |
| | $d_c(q_1) = \frac{h_c(q_1)}{s(c)}$ | R = 0.9999 |
| $(q_1-1)/2$ | $h_c(q_1) = 0.9265 \ln(q_1) - 0.1898$ | R = 0,9966 |
| | $d_c(q_1) = \frac{h_c(q_1)}{s(c)}$ | R = 0.9971 |

Graph 4 below shows $h_c$ and $d_c$ when $j_{q_1,c}(0)$ varies in $[\![0, (q_1-1)/2]\!]$, for $q_1 = 727$. We notice that $h_c\left(j_{727,c}(0)\right)$ and $d_c\left(j_{727,c}(0)\right)$ are within a relative interval of $\pm\,3.5\%$ around their respective average value, equal to 5,8836 and 0,8517%. Similar results were obtained for all studied values of $q_1$.

**Graph 4**: $h_c$ and $d_c$ as a function of $j_{727,c}(0) \leq 363$ [EL3].

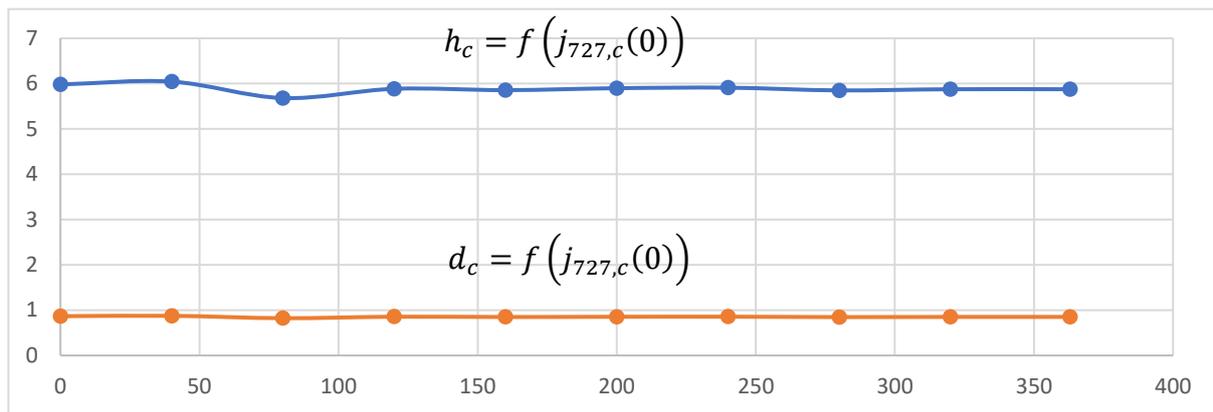

For a fixed value of $q_1$, there exists infinitely many values of $c$ such that $q_1 = \min \mathcal{D}_p(E_c)$, obtained from a finite number (corresponding to $\tilde{\mathcal{C}}_{q_1}$) of congruences modulo $q_1\#$. For $c(0) \in \tilde{\mathcal{C}}_{q_1}$, we then have $c(n) = q_1\# \cdot n + c(0)$ with $m_c(n) = 1 + \log_{10}(c(n))$. The set of possible congruences of $c$ is obtained by algorithm 1. Graph 5 shows, for $q_1 = 727$ and $c(0)$ fixed, how $h_c$ and $d_c$ vary with $m_c(n)$. $c(0)$ was chosen such that $j_{q_1,c}(0) = 0$.

**Graph 5**: Values of $h_c$ and $d_c$ for $q_1 = 727$ as function of $m_c(n)$ [EL3].

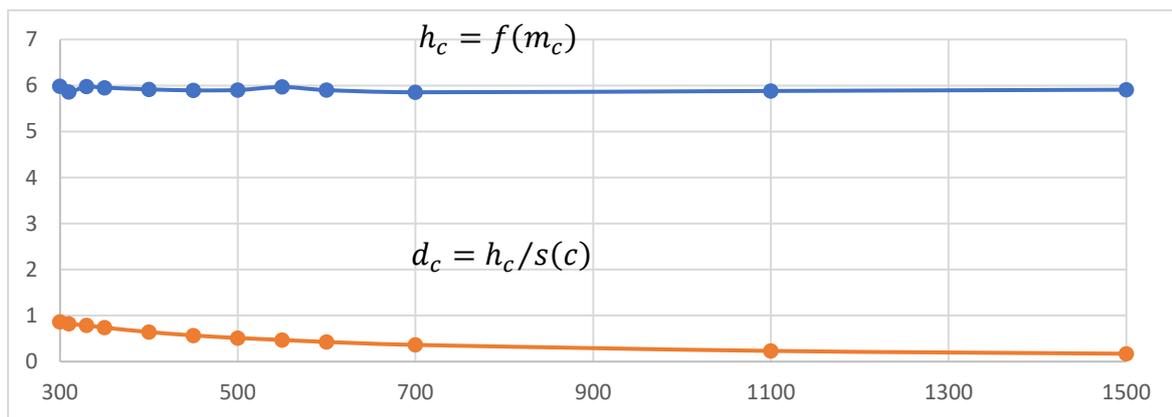

We notice that $h_c$ is in a relative range of $\pm 1.5\%$ around its average value. This weak variation of $h_c = \prod_{p \in \mathbb{P} \setminus \{2\}} \frac{p - t_p(c)}{p-1}$ can be explained by the fact that $t_p(c(n)) = t_p(c(0))$ for any $p \leq q_1$. Although for the primes $p > q_1$ we cannot predict the value of $t_p(c(n))$ we note that their cumulated effect on $h_{c(n)}$ seems weak.

The results of this section show that $h_c$ mainly depends on $q_1$ while parameters $j_{q_1,c}(0)$ and $m_c$ only weakly impact it.

### 5.2.2. Distribution of primes

In this section, we study the distribution of primes in $E_c$ for $X$ varying in $[\![0, 4m_c]\!]$. The purpose is to heuristically predict the minimal range where a prime number can be found. We use the same notations as in the previous section. The results below were obtained with $q_1 + 1$ pairs, more precisely with 2 different values of $c$ per value of $j_{q_1,c}(0)$. Furthermore, to have a constant size for all values of $c$ with $q_1$ fixed, $c$ was always chosen between $q_1\#$ and $2q_1\#$. The pairs studied in this section are in [EL7] (for details see section 4-2 in [EL3]). Numerical series are compiled in [EL3].

**Definition 5.2.2**: Let $I_z = \left[\!\left[0, \frac{4m_c}{z}\right]\!\right], z \in [1, +\infty)$. We let $N_z = |\{X^2 + c; X \in I_z\} \cap \mathbb{P}|$.

**Remark 5.2**: Conjecture 2 gives, for large values of $c$, $N_z \sim \frac{4}{z.ln(10)} h_c$.

We will first show that for all the pairs, the prime numbers of $\{X^2 + c; X \in I\}$ are distributed in similar quantities, for intervals $I$ forming a regular subdivision of $I_1$. Then we will show that $N_z$ is close to its equivalent value given in remark 5.2. Finally, we will show that the proportion of pairs for which there exists at least one prime number in the interval $I_z$ increases with the value of $q_1$.

For $q_1 = 727$, we studied 728 pairs of values verifying $m_c = 301$. The value of $h_c$ estimated in the previous section of 5.9860 allows us to estimate $N_1 \simeq 10.39$. Table 2 gives the number of primes in a regular partition of ten subdivisions, and the average number of primes. We notice that the prime numbers are distributed approximately uniformly in each interval within a relative tolerance of 9.5%.

**Table 2**: Number of primes in each interval $\Delta X = 0.4 m_c$ for $q_1 = 727$ and 728 pairs [EL7].

| Value of $j(X) = \frac{X}{2}$ | Number of primes for the 728 pairs of values | Average number of primes per pair of values |
|---|---|---|
| $[\![0,60]\!]$ | 761 | 1.05 |
| $[\![61,120]\!]$ | 751 | 1.03 |
| $[\![121,180]\!]$ | 755 | 1.04 |
| $[\![181,240]\!]$ | 756 | 1.04 |
| $[\![241,300]\!]$ | 704 | 0.97 |
| $[\![301,360]\!]$ | 721 | 0.99 |
| $[\![361,420]\!]$ | 753 | 1.03 |
| $[\![421,480]\!]$ | 738 | 1.01 |
| $[\![481,540]\!]$ | 741 | 1.02 |
| $[\![541,600]\!]$ | 768 | 1.05 |
| $[\![0,600]\!]$ | 7448 | $10.23 \sim N_1$ |

We define the density of $N_z(n)$ as the percentage of pairs of values for which the number of primes is $n$ in the same interval $I_z$. The graph 6 shows it for $z = 1$ and $z = 2$. For $q_1 = 727$, the densities peak at $\lfloor N_2 \rfloor = 5$ and $\lfloor N_1 \rfloor = 10$. These results generalize experimentally as soon as $q_1 \geq 251$, even when shifting the interval $I_z$.

These results suggest that the density of $N_z$ <u>over all</u> the pairs is regular. The peak of the distribution is located around the equivalent of $N_z$ provided by remark 5.2.

**Graph 6**: densities (%) $N_1$ and $N_2$ for $q_1 = 727$ [EL3].

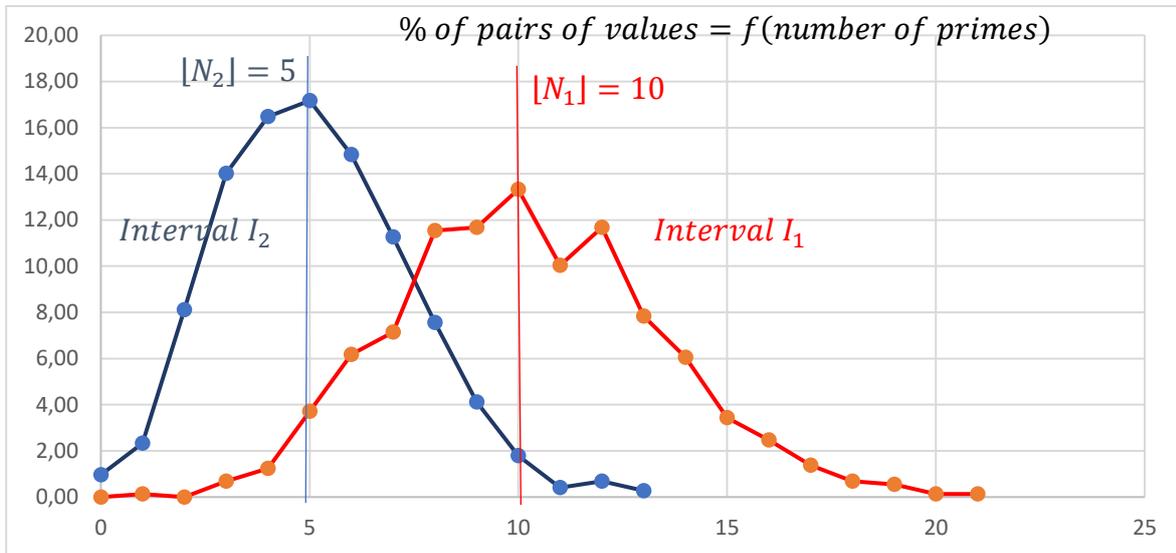

For all pairs studied, there is at least one prime in $I_1$. When $z > 1$, some pairs no longer have one in $I_z$. Graph 7 shows, however, that this tends to disappear when $q_1$ increases. The curves below are obtained for $z \in \left\{1, 2, \frac{10}{3}, 5\right\}$.

**Graph 7**: Empirical probability that $I_z$ has at least one prime, when $q_1$ varies ([EL3] and [EL7]).

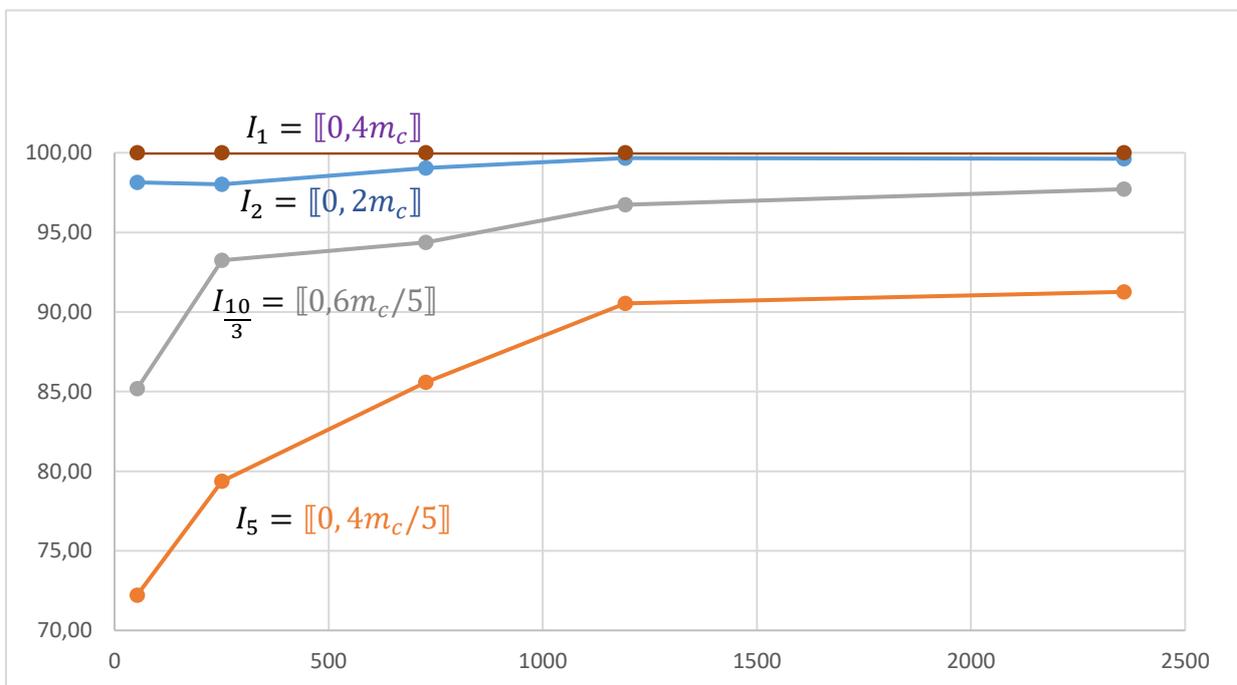

These results suggest that as $q_1$ increases, the search for large primes can be performed in an interval of proportionally smaller size, which can be achieved by selecting a higher value of $z$.

Note: As $q_1$ increases, the estimated value of $h_c$ depends less and less on the values of $n$ and $j_{q_1,c}(0)$ used. For example, for $q_1$ varying from 131 to 1193, the deviation from the mean of the estimates goes from $\pm\ 6.5\%$. to $\pm\ 1.5\%$.

### 5.2.3. Application to cryptography

The main general public cryptography systems use large prime numbers to secure communications and access data through encryption and decryption operations. These operations use a pair of keys generated from two large prime numbers, the binary size of which is given: 2048, 4096, 8192, etc.

Security is based on the existence of many primes and the difficulty of finding the two prime numbers used to generate a pair of keys.

In this article, the method described allows to find many primes of a given size $s$ by calculating values of $c$ with parameters $q_1$ and $m_c$:

- $q_1$ is chosen so that $q_1\#$ has a size comparable to the objective,
- one remainder of $c$ modulo $q_1\#$ is computed,
- the desired size $m_c$ can then be obtained by adding a multiple of $q_1\#$ to that remainder.

For $q_1 = p_M$ fixed, the theoretical number of congruences is $\frac{q_1+1}{2}\prod_{j<M}\frac{p_j-1}{2}$. The number of primes in the corresponding sets $E_c$, denoted by $N(q_1)$, can be approximated by the number of congruences multiplied by the asymptotic density given by the conjecture 2:

$$\frac{h_c}{m_c.\ln(10)}\frac{q_1+1}{2}\prod_{\forall j<M}\frac{p_j-1}{2} \lesssim N(q_1) \lesssim \frac{h_c.10^{\frac{m_c}{2}}}{m_c.\ln(10)}\frac{q_1+1}{2}\prod_{\forall j<M}\frac{p_j-1}{2}$$

This method therefore seems to meet the requirements of cryptography.

Note: The factor $10^{\frac{m_c}{2}}$ is for eliminating redundant primes generated by the method, due to the fact that a prime generally belongs to several sets $E_c$. We will not study collisions further in this article.

In table 3, we present the average computation time to obtain 2 primes of a given size in bits. Algorithm 2 was used to compute $c$ congruences. The search was performed for $z = 4$ using 100 pairs of values $\left(c, j_{q_1,c}(0)\right)$. The calculations were performed on a laptop with an Intel(R) Core(TM) i7-10750H CPU @ 2.60GHz, and parallelized over 10 threads.

**Table 3**: Calculation time per couple of primes found.

| Binary size (bits) | 2048 | 4096 | 8192 |
|---|---:|---:|---:|
| Decimal size $(m_c - 1)$ | 616 | 1233 | 2466 |
| $q_1$ | 1471 | 2887 | 5779 |

| $n$ | 10 | 100 | 10 |
|---|---|---|---|
| Computation time per couple of primes (seconds) | 0.21 | 2.15 | 20.72 |

This method is also applicable to the search for very large primes, i.e. numbers much larger than those used in cryptography. With $q_1 = 115\,631$ and $j_{q_1,c}(0) = 0$, we determined 2 prime numbers with 50 001 digits [EL5] of the form $X^2 + c$ such that $j(X) < m_c/2$. To determine larger primes, the primality tests would have to be spread over a large number of computers or a supercomputer.

We also calculated $c(0)$ for $q_1 = 57\,571\,307$ (respectively $q_1 = 115\,140\,317$) and $j_{q_1,c}(0) \in [\![0,8]\!]$, which should allow obtaining primes with more than 25 million digits (respectively 50 million digits). These values are presented in [EL6]. The number of primes in $I_z$ for $z = 2$, which is estimated with the relation $(H1)$ in table 1, would then be expected around 13 (respectively 14). For one billion digits, we could keep increasing $q_1$ or add $10^{95.10^7}.q_1\#$ to the second congruence. In the second case, the number of primes would still be expected to be around 14 in $I_z$ for $z = 2$.

## 6. CONCLUSION

We have determined a method for obtaining large primes from the study of the density of primes of the form $X^2 + c$. To this end, we presented a method that consists in fixing the smallest prime divisor of $E_c$. Furthermore, with a uniform version of Shanks' conjecture [2] we showed that the asymptotic density $d_{\mathbb{P}|E_c}(\sqrt{c})$ of primes in $E_c \cap [0,2c]$ is given by:

$$d_{\mathbb{P}|E_c}(\sqrt{c}) \underset{c \to +\infty}{\sim} \frac{1}{\ln(10)} \cdot \left(\frac{h_c}{m_c}\right)$$

With the results of [6], we also conjectured that for $p_1 \in \mathcal{D}_p(E_c)$ fixed, there exists an infinity of $p_2$ primes such that $p_1 p_2 \in E_c$.

Finally, we showed that $h_c$ grows logarithmically with $q_1 = \min \mathcal{D}_p(E_c)$, which allows us to search for primes in $E_c$ for smaller and smaller values of $X$ relative to $m_c$. We have thus shown that quadratic forms can be used in cryptography to determine large primes of a fixed size, as well as very large primes.

Acknowledgments: We would like to thank François-Xavier VILLEMIN for his attentive comments and suggestions.

# EXTERNAL LINKS

# APPENDIX A: Maple regression - graph 1

La table A-1 shows the number of primes $\left|\mathbb{P}^{(x)} \cap \mathcal{B}_{5,1,1}\right|$ as function of $x$. The numeric values were obtained using Maple's function « isprime » applied to elements of the quadratic progression $\mathcal{B}_{5,1,1}$.

Table A-1: numeric values of $\left|\mathbb{P}^{(x)} \cap \mathcal{B}_{5,1,1}\right|$.

| $x$ | $1 \times 10^4$ | $5 \times 10^4$ | $1 \times 10^5$ | $5 \times 10^5$ |
|---|---|---|---|---|
| $\left|\mathbb{P}^{(x)} \cap \mathcal{B}_{5,1,1}\right|$ | 482 | 1904 | 3541 | 15 217 |

| $x$ | $1 \times 10^6$ | $5 \times 10^6$ | $1 \times 10^7$ | $2 \times 10^7$ |
|---|---|---|---|---|
| $\left|\mathbb{P}^{(x)} \cap \mathcal{B}_{5,1,1}\right|$ | 28 810 | 128 580 | 245 094 | 468 277 |

| $x$ | $3 \times 10^7$ | $4 \times 10^7$ | $5 \times 10^7$ | $6 \times 10^7$ |
|---|---|---|---|---|
| $\left|\mathbb{P}^{(x)} \cap \mathcal{B}_{5,1,1}\right|$ | 684 782 | 896 539 | 1 106 006 | 1 312 328 |

| $x$ | $7 \times 10^7$ | $8 \times 10^7$ | $9 \times 10^7$ | $1 \times 10^8$ |
|---|---|---|---|---|
| $\left|\mathbb{P}^{(x)} \cap \mathcal{B}_{5,1,1}\right|$ | 1 517 012 | 1 720 556 | 1 922 292 | 2 122 714 |

For the regression of $\left|\mathbb{P}^{(x)} \cap \mathcal{B}_{5,1,1}\right|$, we used Maple's routine **NonlinearFit** with empirical initial values $a$ and $b$ as below:

**NonlinearFit**($a \times n^b$, X, Y, n, initialvalues = [$a$ = .2, $b$ = .9], output = [leastsquaresfunction, residuals])

We get the following regression:

$$\left|\mathbb{P}^{(x)} \cap \mathcal{B}_{5,1,1}\right| \simeq 0.0647540289799160 \times x^{0.939431500805529}$$

$$R = 0.999999677734660$$